\documentclass[aps,pra,twocolumn]{revtex4-2}  
\usepackage{graphicx}%
\usepackage{amsmath,amssymb,amsfonts}%
\usepackage{amsthm}%
\usepackage{xcolor}%
\usepackage{braket}
\usepackage{comment}
\usepackage{mathrsfs}  

\usepackage{subcaption}

\begin{document}

\title[Article Title]{Capturing spin chain dynamics with periodically projected time-dependent basis }

\author{Grace Hsiao-Han Chuang}
\email{hhchuang@pks.mpg.de}
\affiliation{Max-Planck-Institute for the Physics of Complex System,  Nöthnitzer Straße 38, Dresden, 01187, Germany}

\author{Abhijit Pendse}
\affiliation{Max-Planck-Institute for the Physics of Complex System,  Nöthnitzer Straße 38, Dresden, 01187, Germany}

\begin{abstract}
Simulating many-body quantum systems poses significant challenges due to the large size of the state space. To address this issue, we propose using an SU(2) coherent state for individual spins to simulate spins on a lattice and derive equations of motion based on the variational principle. This method involves a sampling approach, where a subset of relevant configurations is chosen based on energy criteria, and a projection method is used to remove linear dependency on the overcomplete and time-dependent basis during propagation. We validate this method through numerical simulations of up to seven-qubit system, calculating key physical observables such as state probabilities and domain-wall densities. Our results indicate that while complete basis sets offer accurate dynamics, selected incomplete sets can recover essential features, especially with the assistance of a projector. The selected incomplete dual bases method is not limited by the structure of Hamiltonian and efficiently captures the non-equilibrium dynamics. 
\end{abstract}

\maketitle

\section{Introduction}\label{sec:intro}
A chain of coupled two-level atoms known as the Ising chain is the simplest form of a many-body quantum system~\cite{ising_lienhard,ising_manybody_browaeys}. An Ising chain with tunable parameters is of interest for studying many-body dynamics, and phase transitions, and for use in quantum technologies. To simulate this system numerically, one needs to solve the time-dependent Schr\"{o}dinger Equation~\cite{Sakurai_1993} by writing the state vector as a linear combination of a set of complete, orthonormal and time-independent basis functions. 
\begin{equation} \label{eqn:basis_notime}
    |\boldsymbol{\Psi}(t)\rangle = \sum^N_{\ell=1} \alpha_\ell(t) |\boldsymbol{\zeta_\ell}\rangle,
\end{equation}
This is the most general way to express the wave function in quantum dynamics. The time-dependence of the state vector is captured by the amplitudes as~\cite{Schweizer_2006}, where $N$ is the size of the basis set $\{|\boldsymbol{\zeta}_\ell\rangle\}$. 

Various types of time-independent basis functions are utilised across different problems. 
In the context of electron transfer, transforming adiabatic states to diabatic\cite{ruedenbergQuantumChemicalDetermination1993,hendekovicNovelVariationalDefinition1982} helps model electron coupling. The chosen degree of freedom is usually 2 to 4 (see Chapter 4~\cite{domckeConicalIntersectionsElectronic2004}). 
Several methods for constructing the transformation matrix to form diabatic states are proposed~\cite{zhangInitioQuantumChemical1997,subotnikConstructingDiabaticStates2008, caveGeneralizationMullikenHushTreatment1996,caveGeneralizationMullikenHushTreatment1996,hammes-schifferProtoncoupledElectronTransfer2008}.

A more general time-independent basis is Discrete Variable Representation (DVR)~\cite{Littlejohn_2002,Light_2007}. The simplest DVR method employs spatial delta functions~\cite{moritaMultidimensionalOHLocal2013}. More advanced DVR techniques have been developed to reduce the cost~\cite{Webb_2000,Baraban_2011}. However, these methods require a complete set of functions and are therefore affected by the exponential growth of system size. Although DVR is widely used, it is generally limited to systems with few degrees of freedom~\cite{Colbert_1992}.

Beyond specific problems like calculating electron coupling and reducing complex interactions to simpler subspaces, the use of a time-independent basis generally suffers from the exponential curse due to the necessity of a complete basis set.
To solve this issue, a time-dependent basis is used broadly in quantum dynamics.

Unlike the time-independent basis set, the time-dependent basis set $\{|\boldsymbol{\xi}_\ell(t)\rangle\}$ does not need to be complete.  The number of basis elements depends on the sampling scheme and the evolution of the basis. This set in general are not orthogonal ($\langle \boldsymbol{\xi}_\ell|\boldsymbol{\xi}_m\rangle \neq 0$), but can be normalised  ($\langle \boldsymbol{\xi}_\ell|\boldsymbol{\xi}_m\rangle = 1$). The resulting wave function is also a linear combination of basis functions, and now both amplitude and basis functions are time-dependent.
\begin{equation}\label{eqn:basis_time}
    |\boldsymbol{\Psi}(t)\rangle = \sum^N_\ell a_\ell(t) |\boldsymbol{\xi}_\ell(t)\rangle
\end{equation}

The widely used time-dependent basis is Coherent State (CS)~\cite{Zheng_1990,Combescure_Robert_2012} which was introduced by Schr\"{o}dinger~\cite{steinerSchrodingersDiscoveryCoherent1988} and generalised in Lie group by Perelomov~\cite{perelomovGeneralizedCoherentStates1986}. The canonical CS (Gaussian CS) is particularly useful in non-adiabatic molecular dynamics~\cite{songCoherentStateApproach2006,freixasNonadiabaticExcitedStateMolecular2021,curchodInitioNonadiabaticQuantum2018,gelinInitioSurfaceHoppingSimulation2021}. Because it satisfies the minimum uncertainty principle over time, the quantum dynamics can be simulated in semiclassical way where one solves the classical equation of motion with frozen Gaussian functions~\cite{hellerFrozenGaussiansVery1981} which represents the molecular wave function. 

In this article, we employ the time-dependent SU(2) coherent state as the basis~\cite{arecchiAtomicCoherentStates1972} and draw inspiration from the molecular dynamics~\cite{Shalashilin_Child_2008} to describe the dynamics of 1D Ising spin chain systems using the variational principle. Due to the time-dependent nature of the basis, a complete basis set is not required, offering the potential to mitigate the exponential curse.

An experimental realisation of such an Ising chain may be done using Rydberg atoms~\cite{ising_lienhard,ising_manybody_browaeys,ising_chain_de,ising_rydberg_review_schauss}. Rydberg atom chains have been used to study many-body aspects of spin chains~\cite{Lukin_2017,KibbleZurek_keesling} as well as its use in metrology~\cite{rydberg_metrology_ding} and quantum computation~\cite{ryd_quant_tech_adams}. Rydberg atoms present a suitable platform due to the control over atom placement achieved using optical tweezers, as well as due to the long-range Rydberg-Rydberg interactions~\cite{3ryd_chain_exp_barredo,trapping_rydberg_wilson}. In a recent work~\cite{Lukin_2017} coupled qubits have been realised experimentally as arrays of trapped Rydberg atoms. 
Such a system is analogous to a 1D Ising spin chain and admits quantum many-body states. By varying the ratio of detuning to the Rabi frequency in such a system, one can go from a disordered phase to a Rydberg crystal phase\cite{Lukin_2017}. We apply our method to study the 1D domain wall density and quench dynamics of this experiment and compare the efficacy of our method against the full quantum dynamics of time-independent states in the complete Hilbert state.
\par

In many-body localised systems, local observables (such as domain walls in quench dynamics) do not thermalise and retain the memory of the initial state, indicating a breakdown of ergodicity. Recent experiments ~\cite{Bloch_2012,labuhnTunableTwodimensionalArrays2016,Lukin_2017,Adler_2025} have observed unexpectedly long-lived revivals in these specific high-energy states. This phenomenon has garnered significant interest among theorists, as it violates the eigenstate thermalisation hypothesis (ETH) ~\cite{PhysRevA.43.2046,PhysRevE.50.888}.

In Section~\ref{sec:method}, we illustrate the methodology of using dual bases with projectors. For a general Hamiltonian, we derive the equation of motion for time-dependent SU(2) coherent states ($\{\ket{\boldsymbol{\xi}(t}\}$) using the variational principle in Section~\ref{subsec:eom}, with further details provided in Supporting Information~\ref{si:eom}. Since the choice of basis is flexible, we sample these subspaces based on energy criteria in Section~\ref{subsec:sampling}, with additional details in Supporting Information~\ref{si:num_basis}. To \textit{optimise} or \textit{prune} the time-dependent basis, we employ a projector in Section~\ref{subsec:dualB}. In Section~\ref{sec:result}, we present the numerical results for two physical observables: state probability and domain-wall density, to validate our method. Finally, we summarise our conclusions in Section~\ref{sec:conclusion}.

\section{Methodology}\label{sec:method}
\subsection{Dynamics with time-dependent SU(2) coherent state}\label{subsec:eom}
As an illustration of a Hamiltonian with spin states on a lattice, we consider a 1D chain of atoms coupled to a Rydberg state. However, the formalism that we introduce to numerically propagate the system is not limited to 1D systems and can be generalised to arbitrary spin systems on a lattice. 
\par 
Consider a 1D chain of $M$ trapped neutral atoms, all in the atomic ground state $\ket{g}$. By using a laser with Rabi frequency $\Omega$, each atom is coupled to a Rydberg state $\ket{r}$. The lasers are detuned by $\Delta$ concerning the $\ket{g}\rightarrow\ket{r}$ transition. Due to the dipole moment of the Rydberg state, the atoms at any two sites $i$ and $j$ may interact via repulsive van der Waals interaction $\displaystyle V_{ij} = \frac{C_{6}}{R_{ij}^6}$. In experiments that trap ground state atoms, the Rydberg state is usually anti-trapped \cite{Rydberg_antitrapped}.  As a result, before coupling the ground state atoms to the Rydberg state, one turns off the trapping potentials. The Hamiltonian of the system is then given by
\begin{equation}\label{eqn:hamiltonian}
	\frac{\hat{H}(t)}{\hbar} = 
	\frac{\Omega(t)}{2}\sum^M_i \hat{\sigma}_x^i
	- \Delta(t) \sum^M_i \hat{n}_i 
	+ \sum^M_i\sum^M_{j \ne i } V_{ij} \hat{n}_i \hat{n}_j,
\end{equation}
where $\hat{\sigma}_x^{i}=\ket{g_{i}}\bra{r_{i}} +\ket{r_{i}}\bra{g_{i}}$ denotes the Pauli spin operator coupling the ground state and Rydberg state for the $i$-th atom and $\hat{n}_{i}=\ket{r_{i}}\bra{r_{i}}$. We have considered the Rabi frequency as well as the detuning of the coupling laser to be spatially uniform, but time-dependent.

For a general Ising model Hamiltonian, we use a coherent state basis to expand the time-dependent wave function, as given in Equation~\eqref{eqn:basis_time}. Various types of coherent states can be formulated depending on the specific purpose~\cite{arecchiAtomicCoherentStates1972}. Inspired by the concept of Gaussian coherent state from molecular dynamics~\cite{Shalashilin_Child_2008}, we express this state as an uncorrelated  SU(2)  coherent state (c.f. eqn. 1.7 in~\cite{arecchiAtomicCoherentStates1972}), which is a tensor product of 1D spin coherent states. The key assumption is that SU(2)  coherent states evolve independently according to the variational principle~\cite{tdvp_1981}. Subsequently, the wave function is constructed using this basis, with the amplitudes determined through the variational principle.

The 1D spin coherent state is 
\begin{equation}
    \ket{\boldsymbol{\xi}_\ell(t)} = \bigotimes^M_i  \ket{\xi^i_\ell(t)}, 
\end{equation}
$M$ is the number of atoms. Recall that to express wave function in Equation~\eqref{eqn:basis_notime} and \eqref{eqn:basis_time}, $N$ is the number of basis functions. If one use complete basis, $N=C_{\mathrm{tot}}$, where $C_{\mathrm{tot}}=2^M$. Each $\ket{\xi^i_\ell(t)}$ represents an SU(2) spin coherent state in the $i^{\text{th}}$ site. 
We define the coherent state as~\footnote{Following the advice of Dr J. Rawlinson, we parametrise Equation~\eqref{eqn:cs1} for the spin coherent state in a way that is numerically more stable than the more common parametrisation, $\displaystyle \ket{\xi}=\frac{\ket{g}+\xi{\ket{r}}}{\sqrt{1+|\xi|^{2}}}$. In the common expression, as $\xi\rightarrow\infty$, the state $\ket{\xi}$ approaches $\ket{r}$, which poses challenges in numerical simulations.}
\begin{equation} \label{eqn:cs1}
    \ket{\boldsymbol{\xi}_\ell(t)} 
        = \bigotimes^M_i  \ket{\xi^i_\ell(t)} 
        = \bigotimes^M_i  
        \frac{
            (\xi^i_\ell(t)+1) \ket{g_i} 
            + (\xi^i_\ell(t)-1) \ket{r_i}}
            { \sqrt{ 2
            \big(
                1+|\xi^i_\ell(t)|^{2}
            \big)}
        },
\end{equation}
where $\ket{g_i}$ and $\ket{r_i}$ are the two eigenstates of the $i^{th}$ two level system or eigenvectors of the Pauli matrix, $\hat{\sigma}_{z}^{i}$.
In Equation~\eqref{eqn:cs1}, $\xi^i_\ell$ is a complex number with magnitude $0 \leq |\xi| \leq 1$, and $\bar{\xi}^i_\ell$ denotes its complex conjugate. This coherent state basis, $\{\ket{\boldsymbol{\xi}}\}$, are normalised ($\bra{\boldsymbol{\xi}_\ell}\boldsymbol{\xi}_\ell\rangle = 1 $) to avoid nonphysical interpretation, but it is not orthogonal ($\bra{\boldsymbol{\xi}_m} \boldsymbol{\xi}_\ell\rangle \ne 0 $ for $m\ne l$). The corresponding overlap matrix is 
\begin{equation} \label{eqn:overlap}
    \Gamma_{ml} \equiv \bra{\boldsymbol{\xi}_m} \boldsymbol{\xi}_\ell\rangle = \prod^M_i 
    \frac{
        1 + \bar{\xi}^i_m \xi^i_\ell
    }{
        \sqrt{ 1 + |\xi^i_m|^{2}}
        \sqrt{ 1 + |\xi^i_\ell|^{2}}
    }.
\end{equation} 

To solve the time-dependent Schr\"{o}dinger equation, we construct the Lagrangian using the wave function from Equation~\eqref{eqn:cs1} and the Hamiltonian from Equation~\eqref{eqn:hamiltonian}. The equation-of-motion is then derived using the time-dependent variational principle~\cite{tdvp_1981}. The complete derivation can be found in the Appendix~\ref{si:eom}.
\begin{align}
    \dot{\vec{a}} &= \mathbf{\Gamma}^{-1}
        (
            -\mathbf{X}- i \; \mathbf{H}
        )  \; \vec{a}, \label{eqn:eom_a} \\
    \dot{\xi}^i_\ell &= -i 
        \Bigg[
            \Omega(t) \xi^i_\ell + \frac{(\xi^i_\ell)^2-1}{2} \nonumber \\
            &\quad \times \Big( 
                -\Delta(t) 
                +  \sum^M_{j \ne i} V_{ij}
                \frac{
                    (\bar{\xi}^j_\ell-1)(\xi^j_\ell-1)
                }{
                    1+ |\xi^j_\ell|^{2}
                } 
            \Big)
        \Bigg]. \label{eqn:eom_z}
\end{align}
Equation~\eqref{eqn:eom_a} describes the evolution of the amplitude vector $\vec{a}$, where the matrix $\mathbf{X}$ depends on the overlap matrix $\mathbf{\Omega}$. The explicit form of the matrix elements of $\mathbf{X}$ is given in Equation~\eqref{eqn:chi}, while the overlap matrix $\mathbf{\Omega}$ is defined in Equation~\eqref{eqn:overlap}. Similarly, the Hamiltonian matrix $\mathbf{H}$, expressed in the coherent state basis, has its matrix elements detailed in Equation~\eqref{eqn:matrix_element_hamiltonian}. The evolution of a single state vector $\ell$ is governed by Equation~\eqref{eqn:eom_z}, where $\ell$ is constructed as a product of individual coherent states. Each coherent state, represented by $\ket{\xi^i_\ell}$, is obtained through the variational principle. Importantly, Equation~\eqref{eqn:eom_z} demonstrates that the basis remains independent of the amplitude, whereas Equation~\eqref{eqn:eom_a} illustrates that the amplitude depends on the basis. Moreover, since $\mathbf{\Omega}$, $\mathbf{X}$, and $\mathbf{H}$ are functions of $\xi$, their evolution is inherently linked to the time-dependent structure of the basis.

Building on this formulation, our simulation follows a sequential approach. We first compute the time-dependent basis by solving Equation~\eqref{eqn:eom_z}, ensuring that the coherent states evolve correctly. Once the basis is determined, we proceed to calculate the amplitudes using Equation~\eqref{eqn:eom_a}, maintaining consistency between the evolving basis and amplitude dynamics. This structured approach ensures numerical stability and accuracy in our wave function propagation. To implement this methodology, we perform numerical simulations using the Python library QuTip~\cite{qutip1_2012,qutip2_2013}, which provides efficient tools for solving time-dependent quantum systems.

\subsection{Sampling using incomplete orthonormal basis} \label{subsec:sampling}

\begin{figure}[!htbp]
    \centering
    \includegraphics[width=0.7\linewidth]{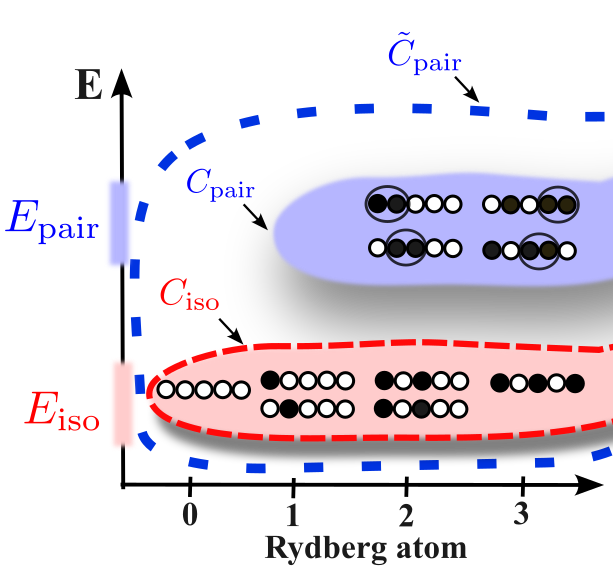}
    \captionsetup{justification=raggedright,singlelinecheck=false} 
    \caption{Illustration of classical configurations for $C_{\mathrm{iso}}$, $C_{\mathrm{pair}}$, and $\tilde{C}_{\mathrm{pair}}$ in a one-dimensional qubit chain. Black circles represent Rydberg atoms, while white circles denote ground-state atoms.}
    \label{fig:classicalConfig}
\end{figure}
The size of the complete basis set is given by $C_{\mathrm{tot}} = 2^M$, where $M$ represents the number of qubits in the chain. However, not all basis functions contribute significantly to the system’s dynamics, so it is necessary to select the most relevant ones for constructing the total wave function. Since the interaction between two Rydberg atoms at sites $i$ and $j$ decreases rapidly as the distance increases ($\displaystyle\frac{1}{R_{ij}^6}$), the strongest interaction occurs between nearest neighbours. As a result, we can define subsets of states based on the \textit{classical configuration}.

Classical configurations describe qubits as being either fully in the ground state or the Rydberg state, forming an orthonormal basis~\cite{Zlatko_2019}. In contrast, coherent states represent each qubit as a linear combination of the ground and Rydberg states, resulting in a non-orthogonal basis.

As illustrated in Figure~\ref{fig:classicalConfig}, we demonstrate the subsets and their corresponding classical configurations. First, we enumerate all possible classical configurations and define a subset ($C_{\mathrm{iso}}$) containing only states with the ground state (no Rydberg atom) and isolated Rydberg atoms. Next, we extend this subset by including states with a single pair of adjacent Rydberg atoms, forming the second subset ($C_{\mathrm{pair}}$). This leads to the definition of an expanded subset, $\tilde{C}_{\mathrm{pair}} = C_{\mathrm{pair}} + C_{\mathrm{iso}}$. Finally, the complete basis set includes all possible configurations ($C_{\mathrm{tot}}$). In our simulation, we use these three sets of basis functions, where their respective sizes are given by:
\begin{equation}\label{eqn:config_num}
    C_{\mathrm{tot}} > \tilde{C}_{\mathrm{pair}} > C_{\mathrm{iso}}
\end{equation}
The analytical expression and detail derivation of these subsets ($C_{\mathrm{tot}}, \tilde{C}_{\mathrm{pair}}$ and $C_{\mathrm{iso}}$) is shown in appendix~\ref{si:num_basis}. 

\subsection{Projector onto basis subsets} \label{subsec:dualB}

The SU(2) Coherent State forms an overcomplete basis set, offering flexibility in constructing the state vector with a minimal number of bases. Equation~\eqref{eqn:cs1} illustrates that by varying the complex numbers $\xi$, a single basis can generate any classical configuration. However, the time-dependent basis $\ket{\boldsymbol{\xi}(t)}$ evolves based on the variational principle, and managing the evolution can be challenging, as we can only control the quantity of bases, not their evolution over time. 

Mathematically, an overcomplete basis set consists of linearly dependent components. During propagation, some basis elements may become redundant or lost. To address this issue, we aim to prune the time-dependent basis by removing redundant components or adding any missing ones. We utilise projectors during propagation to accomplish this goal.

At any snapshot $t'$, the wave function can be written in both time-independent and time-dependent basis sets, 
\begin{equation}
    \ket{\boldsymbol{\Psi}(t')} = \sum^N_{\ell=1} a_\ell(t') \ket{\boldsymbol{\xi}_\ell(t')} 
            = \sum^{N}_{m=1} \alpha_m(t') \ket{\boldsymbol{\zeta}_m},
\end{equation}
where $N$ represents the number of basis functions, which corresponds to either $C_{\mathrm{iso}}$ or $\tilde{C}_{\mathrm{pair}}$ in our simulation. The value of $N$ does not necessarily have to be the same for these bases; however, we kept it consistent to simplify the simulation.
The projector, $\mathbf{P}(t')$, at this snapshot is constructed by calculating the amplitudes, $\alpha_m(t')$, on a time-independent basis. 
\begin{equation}\label{eqn:projector}
    \alpha_m(t') 
            = \sum^N_{\ell=1}  P_{ml}(t') a_\ell(t'), 
\end{equation}
where the projector is defined as 
\begin{equation}
    P_{m\ell}(t') = \bra{\boldsymbol{\zeta}_m} \boldsymbol{\xi}_\ell (t')\rangle  . 
\end{equation}
In our simulation, we apply this projector every $\Delta t = 0.1 ,\mu$s across all scenarios. We tested $\Delta t$ values ranging from 0.1 to 0.6 $\mu$s and found that the optimal value varies depending on the qubit chain length. For consistency, we select a value that satisfies the criteria for all qubit chains in this study.

We begin with an orthonormal basis and evolve the system using the equations of motion for both the time-dependent basis and amplitude (Equations~\eqref{eqn:eom_a} and~\eqref{eqn:eom_z}). During this evolution, the basis remains normalised but becomes non-orthogonal. After propagating for 0.1 $\mu$s, we apply the projector from Equation~\eqref{eqn:projector} to re-express the wave function on an orthonormal basis. This process is repeated iteratively: evolving the system according to the equations of motion, projecting back to an orthonormal basis, and repeating the cycle. This approach enables us to regulate the basis size, denoted by $N$, while still maintaining access to a complete basis set.
\section{Results} \label{sec:result}

\subsection{Time-Independent Hamiltonian: Domain-wall density} \label{subsec:timeIndep_H}

The dynamics of the $\mathbb{Z}_2$ Rydberg crystal after quenching the detuning ($\Delta=0$), is measured by domain-wall density. The time-independent Hamiltonian is  
\begin{equation}\label{eqn:hamiltonian_time_independ}
    \frac{\hat{H}}{\hbar} = 
        \frac{\Omega}{2}\sum^M_i \hat{\sigma}_x^i
        + \sum^M_i\sum^M_{j = i + 1} V_{ij} \hat{n}_i \hat{n}_j.
\end{equation}
Prepare the wave function in the target state $\ket{\mathbb{Z}_2}$ and propagate it with the above Hamiltonian, and then the system oscillates between two ordered states $\ket{\mathbb{Z}_2}$ and $\ket{\mathbb{Z}_2'}$  ($\ket{rgrgrgr...} \rightleftharpoons \ket{grgrgrg...}$) in a short period. 
This oscillatory behaviour can be observed by measuring the domain-wall density~\cite{Lukin_2017}, which we calculated from Equation~\eqref{eqn:domain_wall_cs} in our simulation.

\begin{figure}[!htbp]
    \captionsetup{justification=raggedright,singlelinecheck=false} 
    \includegraphics[width=\linewidth]{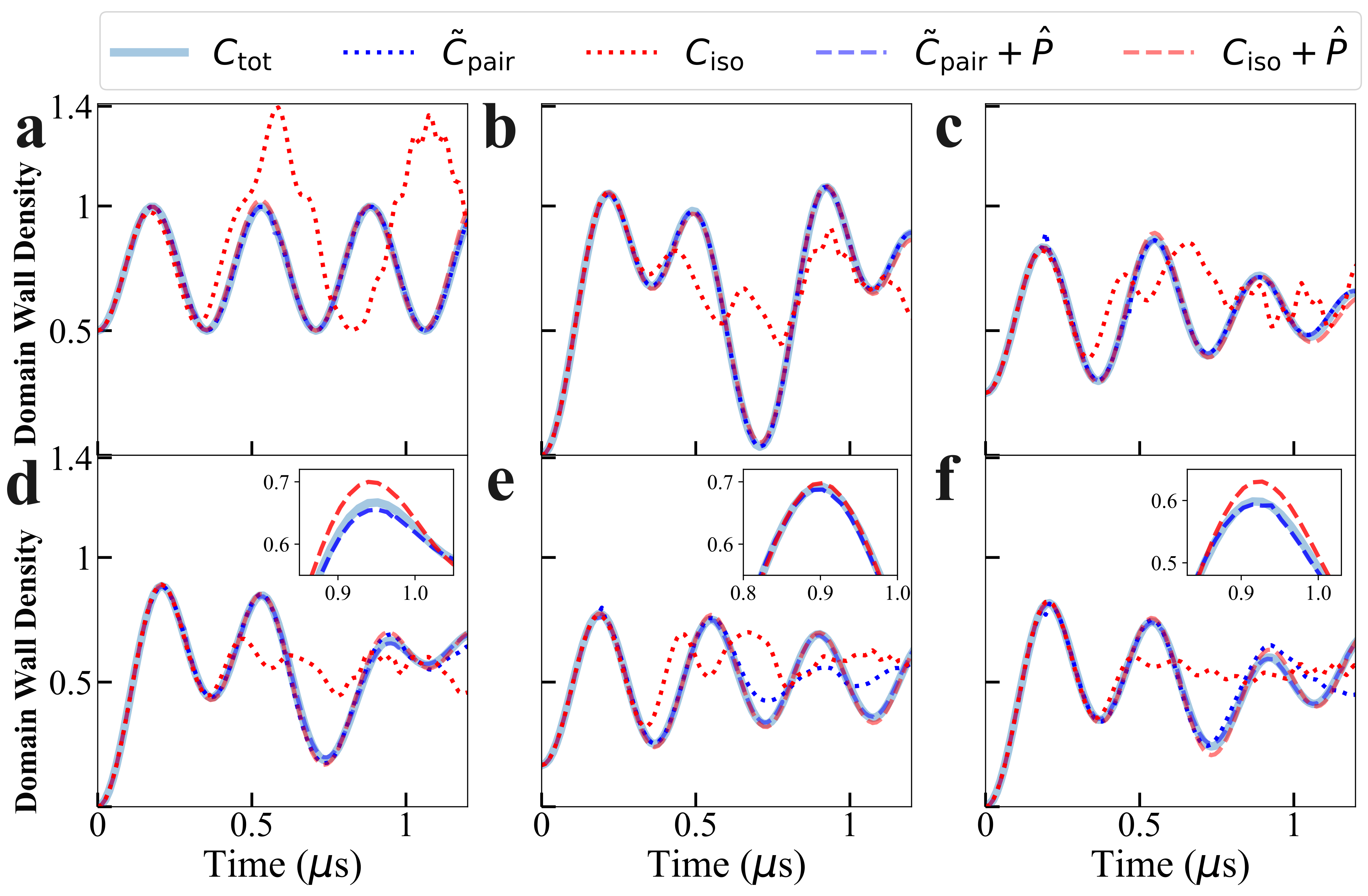}
    \caption{The domain-wall density is presented for various bases of two- to seven-qubit chains, illustrated in subfigures \textbf{a} to \textbf{f}. $C_{\mathrm{tot}}$ uses time-independent complete basis, while $\tilde{C}_{\mathrm{pair}}$ and $C_{\mathrm{iso}}$ use time-dependent incomplete bases. Additionally, $\tilde{C}_{\mathrm{pair}}+\hat{P}$ and $C_{\mathrm{iso}}+\hat{P}$ employ the same bases with projectors.} 
    \label{fig:DW_2to7}
\end{figure}

In Figure~\ref{fig:DW_2to7}, we analyse the domain-wall density across various bases using chains ranging from two to seven qubits. Using the time-independent complete basis ($C_{\mathrm{tot}}$) as a reference, we validate the effectiveness of different time-dependent bases ($\tilde{C}_{\mathrm{pair}}$, $C_{\mathrm{iso}}$, $\tilde{C}_{\mathrm{pair}}+\hat{P}$, $C_{\mathrm{iso}}+\hat{P}$). The number of time-dependent bases considered is $\tilde{C}_{\mathrm{pair}}$ and $C_{\mathrm{iso}}$, where  $\hat{P}$ denotes the involvement of a projector during propagation.

The basis that starts with isolated Rydberg atoms with the $C_{\mathrm{iso}}$ number fails to consistently capture the dynamics from two to seven qubits, indicating that this subset lacks important components.

When we include one pair of Rydberg atoms in the initial configuration ($\tilde{C}_{\mathrm{pair}}$), we find that it improves the accuracy of the domain-wall density for two to five-qubit chains, but it begins to deviate in the long-term dynamics. For example, with the five-qubit chain, the discrepancy in domain-wall density occurs at 1.1 $\mu s$, and for the six-qubit chain, it occurs at 0.6 $\mu s$. This observation suggests that the time-dependent basis ($\{\ket{\xi(t)}\}$ with the initial configuration of $\tilde{C}_{\mathrm{pair}}$) shows limitations in its long-term dynamics.

To capture the non-stationary long-term dynamics for incomplete bases, we "prune" the bases by projecting the time-dependent bases ($\{\ket{\xi(t)}\}$) into the \textit{classical configuration}, either $C_{\mathrm{iso}}$ or $\tilde{C}_{\mathrm{pair}}$. For every 0.1 $\mu s$, we update the wave function by calculating the amplitudes. A detailed explanation of the method can be found in section~\ref{subsec:dualB}. Both projection subsets work effectively to qualitatively restore the non-stationary long-term dynamics. However, projecting to $C_{\mathrm{iso}}$ tends to predict a more disordered state and fewer ordered states. From the inset figures of Figure~\ref{fig:DW_2to7},  $C_{\mathrm{iso}}$ shows a higher domain-wall density at the peak and a lower density at the trough. By including $C_{\mathrm{pair}}$, the order/disorder states achieve better quantitative corrections.

\subsection{Time-dependent Hamiltonian: state population} \label{subsec:timeDep_H}
In accordance with the experiment~\cite{Lukin_2017}, the target state $\ket{\mathbb{Z}_2}$ ($\ket{rgrg…}$) is prepared from the ground state $\ket{\mathbb{G}}$ ($\ket{gggg…}$) by controlling the laser intensities. The time-dependent Hamiltonian, described in Equation~\eqref{eqn:hamiltonian}, is given by:
\begin{equation}\label{eqn:hamiltonian_time_depend}
    \frac{\hat{H}(t)}{\hbar} = 
        \frac{\Omega}{2}\sum^M_i \hat{\sigma}_x^i
        - \Delta(t) \sum^M_i \hat{n}_i 
        + \sum^M_i\sum^M_{j = i + 1} V_{ij} \hat{n}_i \hat{n}_j.
\end{equation}
Here, the Rabi oscillation ($\Omega$) is a constant $2\pi\times2$ MHz within the time range of 0 to 2.25 $\mu$s, and the time-dependent laser detuning ($\Delta(t)$) is represented by the polynomial equation $ \Delta(t) = a + bt + ct^2+dt^3$ , with coefficients $(a,b,c,d) = (-101, 205, -136, 30)$, fitted from 0.6 to 2.3 $\mu$s of experimental data.  The repulsive van der Waals interaction in Rydberg states considers only the nearest neighbour interaction $V_{i,i+1}=2\pi\times 24$ MHz in our model.

\begin{figure}[!htbp]
    \captionsetup{justification=raggedright,singlelinecheck=false} 
    \includegraphics[width=\linewidth]{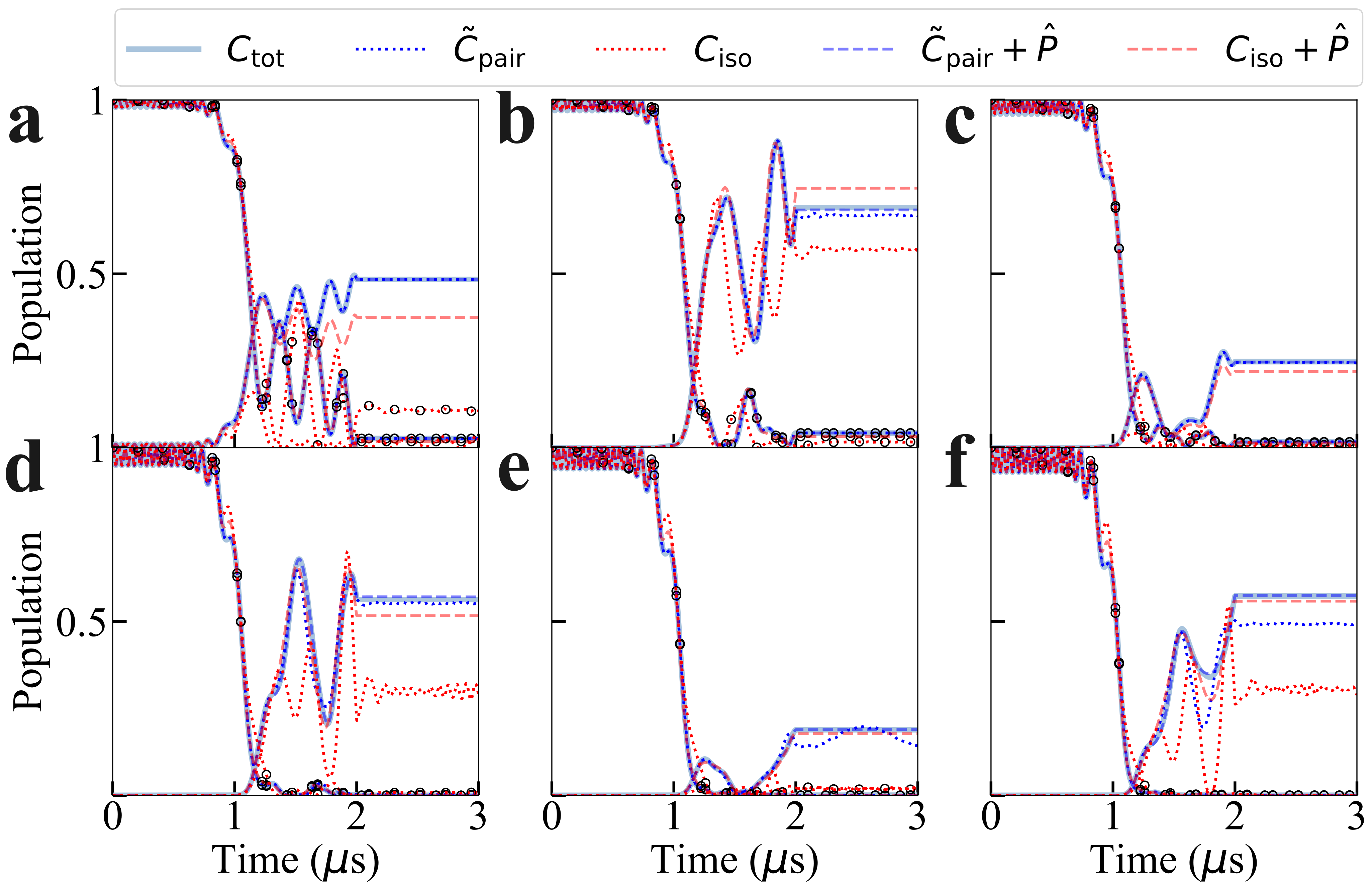}
    \caption{State probability in different bases of two- to seven-qubit chains as illustrated in \textbf{a} to \textbf{f}. Two states are selected: the ground state  $\ket{\mathbb{G}}$ and the target state $\ket{\mathbb{Z}_2}$, with the ground state marked by open circles. $C_{\mathrm{tot}}$ uses time-independent complete basis, while $\tilde{C}_{\mathrm{pair}}$ and $C_{\mathrm{iso}}$ use time-dependent incomplete bases. Additionally, $\tilde{C}_{\mathrm{pair}}+\hat{P}$ and $C_{\mathrm{iso}}+\hat{P}$ employ the same bases with projectors.}
    \label{fig:StateProb_2to7}
\end{figure}

\begin{table}[!h] 
    \captionsetup{justification=raggedright,singlelinecheck=false} 
    \caption{$\ket{\mathbb{Z}_2}$ population of two- to seven-qubit chains  (for $M$ ranging from 2 to 7) at 3 $\mu$s in different bases, with relative deviations (in parentheses) from $C_{\mathrm{tot}}$. $C_{\mathrm{tot}}$ uses time-independent complete basis, while $\tilde{C}_{\mathrm{pair}}$ and $C_{\mathrm{iso}}$ use time-dependent incomplete bases. Additionally, $\tilde{C}_{\mathrm{pair}}+\hat{P}$ and $C_{\mathrm{iso}}+\hat{P}$ employ the same bases with projectors.}
    \centering
    \begin{tabular}{c c c c c c} 
         \hline
          M & $C_{\mathrm{tot}}$ & $\tilde{C}_{\mathrm{pair}}$ & $C_{\mathrm{iso}}$ & $\tilde{C}_{\mathrm{pair}}+\hat{P}$ & $C_{\mathrm{iso}}+\hat{P}$\\ [0.5ex] 
         \hline\hline
         2 & 48\% & 48\% (-0.01\%) & 2\% (-96\%) & 48\% ($\approx 0$\%) & 37\% (-23\%)  \\
         3 & 69\% & 67\% (-3\%) & 57\% (-17\%) & 68\% (-0.9\%) & 75\% (9\%) \\
         4 & 25\% & 24\% (-2\%) & 0.4\% (-98\%) & 25\% (-0.5\%) & 22\% (-12\%) \\
         5 & 56\% & 55\% (-2\%) & 29\% (-49\%) & 57\% (1\%) & 52\% (-8\%) \\
         6 & 19\% & 14\% (-23\%) & 2\% (-89\%) & 19\% (0.3\%) & 18\% (-6\%) \\
         7 & 57\% & 49\% (-14\%) & 29\% (-49\%) & 57\% (0.2\%) & 56\% (-2\%)\\ [1ex] 
         \hline
    \end{tabular}
    \label{tab:Z2popu}
\end{table}

Figure~\ref{fig:StateProb_2to7} illustrates the state probabilities for qubit chains consisting of two to seven qubits. For the seven-qubit chain, we compare our simulation results with the state probabilities obtained from the experiment conducted by Lukin et al.~\cite{Lukin_2017}. Our simulations, whether performed using a time-dependent or time-independent complete basis with a basis size $C_{\mathrm{tot}}=2^M$, where $M$ represents the number of qubits, show strong agreement with the experimental data. For the two- to six-qubit chains, we adopt the time-independent complete basis as the reference. The reference wave function is computed using the time-independent basis defined in Equation~\eqref{eqn:basis_notime}, where this complete basis corresponds to the \textit{classical configuration}. 

In the seven-qubit chain experiment described in~\cite{Lukin_2017}, the system initially has over 80\% probability of being in the ground state $\ket{\mathbb{G}}$. During the short-term evolution (0–1 $\mu s$), the system exhibits consistent oscillations, followed by a rapid decline in the ground state probability. In the long-term assessment (1.5–3 $\mu s$), the probability is primarily influenced by the target state $\ket{\mathbb{Z}_{2}}$, with experimental measurements showing a final $\ket{\mathbb{Z}_{2}}$ ranges from 54\% to 77\%.

In Figure~\ref{fig:StateProb_2to7} \textbf{f}, we present our simulation results for a seven-qubit chain, where we start from the ground state $\ket{\mathbb{G}}$ and propagate the wave function using Equation~\eqref{eqn:hamiltonian_time_independ} in the time-independent complete basis as a reference, and Equation~\eqref{eqn:hamiltonian_time_depend} for various basis choices. Since the precise initial conditions from the experiment are not available, we assume an initial ground state probability of 100\%.

Our model, which considers only nearest-neighbour interactions, predicts a final $\ket{\mathbb{Z}{2}}$ probability of 57\% when using the time-independent complete basis, which aligns well with the experimental range.

There are two key differences between our simulation and the experimental setup. First, we assume an initial ground state probability of 100\%, as the exact initial conditions are unavailable.  Second, our model incorporates only nearest-neighbour interactions, which slightly alters the dynamics compared to the experiment, resulting in a lower final probability for the target state.

To determine the optimal methodology and subspace size for an efficient wave function representation, we explore different time-dependent basis sets with or without projectors, ($\tilde{C}_{\mathrm{pair}}$, $C_{\mathrm{iso}}$, $\tilde{C}_{\mathrm{pair}}+\hat{P}$, $C_{\mathrm{iso}}+\hat{P}$). The number of time-dependent bases considered is $\tilde{C}_{\mathrm{pair}}$ and $C_{\mathrm{iso}}$, where  $\hat{P}$ denotes the involvement of a projector during propagation.

$C_{\mathrm{iso}}$ struggles to capture long-term dynamics, leading to large deviations from  $C_{\mathrm{tot}}$  across all qubit chain lengths as shown Figure~\ref{fig:StateProb_2to7}. For example, in the two-qubit case (Figure~\ref{fig:StateProb_2to7} \textbf{a}, first row of Table~\ref{tab:Z2popu}),  $C_{\mathrm{iso}}$  initially tracks the reference dynamics within 1.0 $\mu$s, but by 3.0 $\mu$s, it severely underestimates the  $\ket{\mathbb{Z}_2}$  population (2\% instead of 48\%).

A similar observation for time-independent Hamiltonian analysis from Section~\ref{subsec:timeIndep_H} indicates that $C_{\mathrm{iso}}$ can only capture short-term dynamics, and it qualitatively and quantitatively fails at capturing long-term dynamics across various qubit chain lengths. 

The same size of single-pair Rydberg basis ( $\tilde{C}_{\mathrm{pair}}$ ) improves accuracy, successfully reproducing short-term dynamics. However, as Table~\ref{tab:Z2popu} shows, its performance deteriorates for larger qubit chains, particularly for  $M = 7$ , where it underestimates the final  $\ket{\mathbb{Z}_2}$  population by 14\%.

Applying a projection method to the incomplete bases during time evolution (Section~\ref{subsec:dualB}) significantly enhances their accuracy. With  $C_{\mathrm{iso}}+\hat{P}$ , we qualitatively recover the expected population trends, while  $\tilde{C}_{\mathrm{pair}}+ \hat{P}$  quantitatively corrects the final  $\ket{\mathbb{Z}_2}$   probabilities, as shown in Table~\ref{tab:Z2popu}. These results demonstrate that projectors effectively mitigate the limitations of incomplete basis sets, allowing for efficient simulation of quantum many-body dynamics without requiring a full basis expansion.

We demonstrate that using a projector to 'prune' the time-dependent basis is effective, regardless of the Hamiltonian's structure with various number of qubit chains. The interaction term in the Hamiltonian allows us to utilise classical configurations to generate important sub-bases, specifically $C_{\mathrm{iso}}$ and $\tilde{C}_{\mathrm{pair}}$. This approach applies to both time-independent Hamiltonians (as discussed in Section~\ref{subsec:timeIndep_H}) and time-dependent Hamiltonians (as discussed in Section~\ref{subsec:timeDep_H}).

\section{Conclusion} \label{sec:conclusion}

Our study demonstrates the effectiveness of using time-dependent SU(2) coherent states combined with a projection-based approach to efficiently simulate the dynamics of quantum many-body systems. By selecting incomplete basis sets based on energy criteria and employing projectors to mitigate redundancy, we successfully captured key physical observables such as state probabilities and domain-wall densities. 

A significant insight from our results is that while a complete basis set provides the most accurate representation of system dynamics, a carefully chosen subset with projectors can recover essential features while drastically reducing computational complexity. This suggests that our methodology can be extended beyond the Ising chain model to other complex quantum systems, including those with longer-range interactions or different coupling structures.

Moreover, our approach offers a practical framework for overcoming the exponential curse of dimensionality in quantum simulations. The ability to extract relevant subspaces dynamically while maintaining numerical stability paves the way for more scalable simulations in quantum computing and condensed matter physics. Future work could focus on extending the method to incorporate next-nearest-neighbour interactions and exploring its applicability to larger systems or experimental implementations.

By leveraging coherent states and projection methods, we provide a promising alternative for studying quantum dynamics without the computational overhead of a full basis set, offering new directions for efficient quantum many-body simulations.

\section*{Acknowledgments}
We thank Dr Dmitry Shalashilin for the discussion and express our deep gratitude to Dr Francisco Gonzalez-Montoya and Dr Jonathan Rawlinson for their invaluable support and profound insights into the equation of motion on a time-dependent basis. We are also deeply grateful to our colleagues, Dr Ritesh Pant and Dr Meng Zeng, for their thorough peer review and constructive feedback. Additionally, we acknowledge the support of the MPI-PKS visitors’ programme. GC also thanks EPSRC Grant No. EP/P021123/1 for the financial support provided.








\bibliography{ref.bib}

\onecolumngrid

\section*{Supplementary information}
\appendix
\setcounter{equation}{0}
\renewcommand{\theequation}{\Alph{section}.\arabic{equation}}

\section{Derivation of Equation~\eqref{eqn:eom_a} and \eqref{eqn:eom_z}}\label{si:eom}

The axiom we use states that the time-dependent bases evolve independently according to variational principle and then the total wavefunction is constructed by these bases weighted by time-dependent amplitude. From equation~\eqref{eqn:cs1}, a single time-dependent basis is represented as a tensor product of one-dimensional SU(2) coherent states. Its Lagrangian is 
\begin{equation} \label{eqn:lagrange_cs}
    \mathscr{L} = \bra{ \boldsymbol{\xi}_\ell } 
        i \frac{
            \hat{ \partial }
        }{
            \partial t
        } 
        \ket{\boldsymbol{\xi}_\ell} - \bra{ \boldsymbol{\xi}_\ell}
        \hat{H} \ket{ \boldsymbol{\xi}_\ell } 
        =i\sum^M_i\frac{\overline{\xi}^i_\ell\dot{\xi}_\ell^i-\dot{\overline{\xi}}_\ell^i\xi_\ell^i}{2(1+|\xi_\ell^i|^2)}
        -\bra{ \boldsymbol{\xi}_\ell}\hat{H}\ket{ \boldsymbol{\xi}_\ell},
\end{equation}
where we use the relation 
\begin{equation}\label{eqn:derPsi}
    \dot{\ket{\boldsymbol{\xi}}}=
    \partial_t\bigotimes^M_i\ket{\xi^i}=\sum^M_i\ket{\xi^i}\bigotimes^{M-1}_{j\ne i}\ket{\xi^j}
\end{equation}
in derivation. 

Using the the Euler-Lagrangian equation we have 
\begin{equation} 
    \frac{d}{dt} \Big(
            \frac{
                \partial \mathscr{L}
            }{
                \partial \dot{ \overline{\xi}^i_\ell }
            } 
        \Big)
        - \frac{
            \partial \mathscr{L}
        }{
            \partial \overline{\xi}^i_\ell
        } = 0, 
\end{equation}
and then the resulting equation of motion for independent coherent state  is 
\begin{equation} \label{eqn:app_eom_z}
    \dot{\xi}^i_l = -i \; \frac{
        \partial \bra{\boldsymbol{\xi}_l} \hat{H} \ket{\boldsymbol{\xi}_l}
    }{
        \partial \overline{\xi}^i_l
    } \; (
        1 + \overline{\xi}^i_l \xi^i_l
    )^2 .
\end{equation}
Each bases is constructed by variational principle independently. Under this axiom, we express the evolution of wavefunction as a linear combination of above time-dependent bases weighted by time-dependent amplitude in equation~\eqref{eqn:basis_time}. To derive the equation of motion of amplitude, we construct the Euler-Lagrangian equation for total wavefunction as 
\begin{equation} \label{eqn:define_lagrange}
    \mathscr{L} = \bra{ \boldsymbol{\Psi} } 
        i \frac{
            \hat{ \partial }
        }{
            \partial t
        } 
        \ket{ \boldsymbol{\Psi} } - \bra{ \boldsymbol{\Psi} }
        \hat{H} \ket{ \boldsymbol{\Psi} }.
\end{equation}
The first term of equation~\eqref{eqn:define_lagrange} is formed by substituting equation~\eqref{eqn:basis_time} and~\eqref{eqn:cs1} in it,
\begin{equation} \label{eqn:1st_lagrange}
    \bra{ \Psi } i \frac{
        \hat{ \partial }
    }{
        \partial t
    } \ket{ \Psi } = 
    i \sum^N_{m,l} \overline{a}_m \dot{a}_l \bra{\boldsymbol{\xi}_m} \boldsymbol{\xi}_l \rangle
    + \overline{a}_m a_l \bra{\boldsymbol{\xi}_m} \dot{\boldsymbol{\xi}}_l \rangle,
\end{equation}
where the first term has an overlap matrix element, $\bra{\boldsymbol{\xi}_m}\boldsymbol{\xi}_l \rangle = \Gamma_{ml}$ and its explicit form is shown in equation~\eqref{eqn:overlap}. Using the relation from equation~\eqref{eqn:derPsi} again, the second term of equation~\eqref{eqn:1st_lagrange}, $ \bra{\boldsymbol{\xi}_m} \dot{\boldsymbol{\xi}}_l \rangle = \chi_{ml}$, which is 
\begin{equation}\label{eqn:chi}
    \chi_{ml} :=  M_{ml} \Gamma_{ml} ,
\end{equation}
where $M_{ml}$ is 
\begin{equation}\label{eqn:Mml_definition}
    M_{ml}:=\sum^M_i 
        \frac{
            \overline{\xi}^i_m \dot{\xi}^i_l
        }{
            1 + \overline{\xi}^i_m \xi^i_l
        }
        - \frac{
            \dot{\overline{\xi}}^i_l \xi^i_l + \overline{\xi}^i_l \dot{\xi}^i_l
        }{
            2( 1 + |\xi^i_l|^2)
        }
\end{equation}
The second term of equation~\eqref{eqn:define_lagrange} is simply  
\begin{equation} \label{eqn:2nd_lagrange}
     \bra{ \boldsymbol{\Psi} } \hat{H} \ket{ \boldsymbol{\Psi} } = 
         \sum^N_{m,l} \overline{a}_m a_l 
            \bra{\boldsymbol{\xi}_m} \hat{H} \ket{\boldsymbol{\xi}_l}   
        =  \sum^N_{m,l} \overline{a}_m a_l H_{ml}. 
\end{equation}
Plug equation~\eqref{eqn:chi} in equation~\eqref{eqn:1st_lagrange}, and then combine with equation~\eqref{eqn:2nd_lagrange}, the explicit form of Lagrangian of equation~\eqref{eqn:define_lagrange} becomes 
\begin{equation} \label{eqn:lagrange}
    \mathscr{L} = \sum^N_{m,l} 
            i \overline{a}_m \dot{a}_l\Gamma_{ml}
        +   i \overline{a}_m a_l \chi_{ml}
        -   \overline{a}_m a_l H_{ml}.
\end{equation}
From the above equations, the equation of motion of amplitude ($a(t)$) is 
\begin{equation} 
    \frac{d}{dt} \Big(
            \frac{
                \partial \mathscr{L}
            }{
                \partial \dot{ \overline{a} }_m
            } 
        \Big)
        - \frac{
            \partial \mathscr{L}
        }{
            \partial \overline{a}_m
        } = 0 .
\end{equation}
Since the equation~\eqref{eqn:lagrange} is not a function of $\dot{\overline{a}}_m$, the first term of the above equation is zero. From the second term $\displaystyle\frac{\partial\mathscr{L}}{\partial\overline{a}_m}=0$, the resulting the equation of motion of amplitude is then
\begin{equation}
    \frac{\partial\mathscr{L}}{\partial\overline{a}_m} 
        = \sum^N_{m,l} i \; \Gamma_{ml} \dot{a}_l 
            + \; i \chi_{ml} a_l - H_{ml} a_l = 0 .
\end{equation}

Rewrite the above equation into matrix form, which is  
\begin{equation} \label{eqn:app_eom_a}
    \dot{\vec{a}} = \mathbf{\Gamma}^{-1}(
        - \boldsymbol{\chi} -i \; \mathbf{H}
        ) \; \vec{a} .
\end{equation}


From equations~\eqref{eqn:app_eom_a} and~\eqref{eqn:app_eom_z}, we derive the equations of motion for both the time-dependent basis and amplitude. The only unknown terms now are $H_{ml}$ and $\displaystyle \frac{\partial \bra{\boldsymbol{\xi}_l} \hat{H} \ket{\boldsymbol{\xi}_l}}{\partial \overline{\xi}^i_l}$. From Ref~\cite{Lukin_2017}, the time-dependent Hamiltonian is given in equation~\eqref{eqn:hamiltonian}. 
Here we present a general form by considering all the interaction between neighbours, and time-dependent laser. After plugin the equation~\eqref{eqn:cs1}, $H_{ml}$ is 
\begin{align} \label{eqn:matrix_element_hamiltonian}
    H_{ml} =\bra{\boldsymbol{\xi}_m} \hat{H} \ket{\boldsymbol{\xi}_l} &= 
    \Bigg(
        \frac{\Omega(t)}{2} \sum^M_i \frac{
            \overline{\xi}^i_m \xi^i_m -1
        }{
            1 + \overline{\xi}^i_m \xi^i_l 
        } 
        - \frac{\Delta(t)}{2} \sum^M_i \frac{
            (
                \overline{\xi}^i_m -1
            )(
                \xi^i_l -1 
            )
        }{
            1+ \overline{\xi}^i_m \xi^i_l 
        }  \nonumber \\
        &\quad +  \sum^M_{j \ne i} \frac{V_{ij}}{4}
        \frac{
            (
                \overline{\xi}^i_m -1
            )(
                \xi^i_l -1
            )
        }{
            1+ \overline{\xi}^i_m \xi^i_l 
        }
        \frac{
            (
                \overline{\xi}^j_m -1
            )(
                \xi^j_l -1
            )
        }{
            1+ \overline{\xi}^j_m \xi^j_l 
        }
    \Bigg) \Gamma_{ml} . 
\end{align}
Replace dummy index $m$ to $l$ and then take derivative with respective to $\overline{\xi}^i_l$, 
\begin{equation} \label{eqn:unknown_in_eom_z}
    \frac{
        \partial \bra{\boldsymbol{\xi}_l} \hat{H} \ket{\boldsymbol{\xi}_l} 
    }{
        \partial \overline{\xi}^i_l
    } = \frac{1}{(1 + |\xi^i_l|^2)^2}
    \Bigg(
        \Omega(t) \xi^i_l + \frac{(\xi^i_l)^2-1}{2}
        \Big(
            -\Delta(t) +  \sum^M_{j \ne i} V_{ij}
            \frac{
                (\overline{\xi}^j_l-1)(\xi^j_l-1)
            }{
                1 + |\xi^j_l|^2
            }
        \Big)
    \Bigg).
\end{equation}

Finally, plugin equation~\eqref{eqn:matrix_element_hamiltonian} into \eqref{eqn:app_eom_a} and equation~\eqref{eqn:unknown_in_eom_z} in \eqref{eqn:app_eom_z}, after that, the explicit form of the equations of motion for both amplitudes $a_l(t)$ and coherent states $\ket{\xi_l(t)}$ are 
\begin{equation}
    \begin{aligned}
        \dot{\vec{a}} &= \mathbf{\Gamma}^{-1}(
            - \boldsymbol{\chi} -i \; \mathbf{H}
            ) \; \vec{a} , \label{eqn:apppen_eom_a}\\
        \dot{\xi}^i_l &= -i \Bigg(
            \Omega(t) \xi^i_l + \frac{(\xi^i_l)^2-1}{2} \Big(
                - \Delta(t) +  \sum ^M_{j \ne i} V_{ij}
                    \frac{
                        (\overline{\xi}^j_l - 1 )(\xi^j_l -1)
                    }{
                        1+ |\xi^j_l|^2
                    }
            \Big)
        \Bigg).
    \end{aligned}
\end{equation}
Equation~\eqref{eqn:apppen_eom_a} is written in the vector-matrix form, and the explicit form of three matrices, and $\mathbf{\Gamma}$, $\boldsymbol{\chi}$, and $\mathbf{H}$, are shown in equation~\eqref{eqn:overlap}, ~\eqref{eqn:chi} and ~\eqref{eqn:matrix_element_hamiltonian}, respectively. 

\section{Derivation of Equation~\eqref{eqn:config_num}} \label{si:num_basis}

We list all the possible micro-configuration under selected conditions to derive the sum of configuration. For example, the total configuration of qubit chain is $2^M$ where $M$ is number of qubit. To consider the micro-configuration, let the number of Rydberg state is $y$, and the number of ground state be $M-y$. If there is no Rydberg state in a chain, there is only one configuration $\ket{gggg\cdots}$, where all qubits are in ground state. If there is one and only one Rydberg state in a chain, the number of micro-configuration is $M$. List all the possible conditions as following equation,
\begin{equation}
    C_{\text{tot}}=\binom{M}{0} + \binom{M}{1} \cdots + \binom{M}{y}+ \cdots +\binom{M}{M} =\sum^M_{y=0}\frac{M!}{(M-y)!y!} = 2^M
\end{equation}

We now consider a chain with isolated Rydberg states, meaning there are no adjacent Rydberg states present. Let $O$ represents the number of dimer configurations  ( $\ket{rg}$ or $\ket{gr}$), which depends on the total number of sites in the chain, denoted as $M$. 
The maximum number of dimer is given by the formula $O=M//2 + 1$,  where $M//2$ represents the integer quotient of $M$ divides by $2$. If the number of ground state configurations, which is ($M-y$), is equal to the number of Rydberg state ($y$), then it follows that $O=M/2=M//2$. Thus, the condition for $O$ can be expressed as 
\begin{equation}
    O = \left\{ 
    \begin{array}{rcl}
         M//2       & \mbox{for} &  M \text{ is even} \\ 
         M//2 + 1   & \mbox{for} & M \text{ is odd},\\
    \end{array}\right.
\end{equation}
The vacancy in the ground state is given by $M-y+1$, considering the two ends in a chain. Place the dimer into the available vacancy, and the resulting configuration of the isolated Rydberg state in the chain is
\begin{align}
    C_{\text{iso}}&=\binom{M-y+1}{0} + \binom{M-y+1}{1} \cdots + \binom{M-y+1}{y}+ \cdots +\binom{M-y+1}{O}  \nonumber  \\
    &=\sum^O_{y=0}\frac{(M-y+1)!}{(M-2y+1)!y!}
\end{align}

We focus on a single pair of Rydberg states, denoted as  $\ket{rr}$, within a chain. Excluding this pair, the remaining number of states in the chain is $M-2$, allowing for a maximum of $\displaystyle\frac{M-2}{2}$ Rydberg states in this section. Consequently, the total maximum number of Rydberg states in the entire chain is given by the formula: $\displaystyle P = (M-2)//2 + 2$. First, we place the pair of Rydberg states in the chain, which can be configured in $M-y+1$ ways. Next, we place the remaining Rydberg states in the available vacancies, which total $M-y$. The overall number of configurations is
\begin{align}
    C_{\text{pair}}&=(M-y+1) \times \Big(
        \binom{M-y}{0} + \binom{M-y}{1} \cdots + \binom{M-y}{y-2}+ \cdots +\binom{M-y}{P} 
        \Big)  \nonumber \\
    &=\sum^P_{y=2}\frac{(M-y+1)!}{(M-2y+2)!(y-2)!}
\end{align}

\section{Domain wall density in time-dependent SU(2) CS}
From the Ref~\cite{Lukin_2017}, the definition of domain-wall density operator per qubit is 
\begin{equation} \label{eqn:dw_eqn}
    \hat{D}=\frac{1}{M} \sum^{M-1}_{i=1}\Big(
        \hat{n}_i\hat{n}_{i+1}+(1-\hat{n}_i)(1-\hat{n}_{i+1})
    \Big) + (1-\hat{n}_1)+(1-\hat{n}_M)
\end{equation}
We calculate the expectation value of $\langle \hat{D} \rangle$ and then $\langle \hat{D}^2 \rangle$ in the following.

The wavefunction in the time-dependent SU(2) Coherent State is defined in equation~\eqref{eqn:basis_time} and \eqref{eqn:cs1}, such that the $\langle \hat{D} \rangle$ is 
\begin{align}\label{eqn:D_expect}
    &\bra{\Psi}\hat{D}\ket{\Psi} \nonumber \\
    &= \frac{1}{M} \sum^N_{m,\ell=1}\bar{a}_m a_\ell
    \bra{\xi_m} \hat{D} \ket{\xi_\ell} \nonumber \\
    &= \frac{1}{M} \sum^N_{m,\ell=1}\bar{a}_m a_\ell
    \Big\{\Big(
        \sum^{M-1}_{i=1} 
        2 \bra{\xi_m}\hat{n}_i \hat{n}_{i+1}\ket{\xi_\ell}
        - \bra{\xi_m}\hat{n}_i\ket{\xi_\ell}
        - \bra{\xi_m}\hat{n}_{i+1}\ket{\xi_\ell}
        \Big)\nonumber \\ 
    & \quad -\bra{\xi_m}\hat{n}_1\ket{\xi_\ell} - \bra{\xi_m}\hat{n}_M\ket{\xi_\ell} + (M+1) \bra{\xi_m}\xi_\ell\rangle 
    \Big\}
\end{align}

To calculate the expectation value, we first calculate the one-body and two-body terms, $\bra{\xi_m}\hat{n}_i\ket{\xi_\ell}$ and $\bra{\xi_m}\hat{n}_i\hat{n}_{i+1}\ket{\xi_\ell}$, and then insert them into the full expression. One-body term is relatively easier, which is 
\begin{equation}\label{eqn:oneB}
    \sum^{M-1}_{i=1}\bra{\xi_m}\hat{n}_i\ket{\xi_\ell} 
    = \frac{\Gamma_{m\ell}}{2}\sum^{M-1}_{i=1} f^i_{m\ell},
\end{equation}
where $\Gamma_{m\ell}$ is the overlap matrix and defined in equation~\eqref{eqn:overlap}. Since $\hat{n}_i = \hat{I}_1\otimes \hat{I}_2\otimes \cdots \hat{n}_i \otimes \hat{I}_{i+1} \cdots$ we can easily get the overlap matrix but site $i$. To simplify the expression, we define a new variable $f^i_{m\ell}$ as 
\begin{equation}
    f^i_{m\ell} \equiv \frac{(\bar{\xi}^i_m-1)(\bar{\xi}^i_\ell-1)}{\bar{\xi}^i_m\xi^i_\ell+1} 
\end{equation}

Same derivation applies on the two-body term is 
\begin{equation}\label{eqn:twoB}
    \sum^{M-1}_{i=1}\bra{\xi_m}\hat{n}_i\hat{n}_{i+1}\ket{\xi_\ell}
    =\frac{\Gamma_{m\ell}}{4}\sum^{M-1}_{i=1}
    f^i_{m\ell} f^{i+1}_{m\ell}
\end{equation}

Insert equation~\eqref{eqn:oneB} and \eqref{eqn:twoB} into equation~\eqref{eqn:D_expect}, 
\begin{equation} \label{eqn:D1}
    \langle \hat{D} \rangle = \frac{1}{M}\sum^N_{m,\ell=1} 
    \bar{a}_m a_\ell \frac{\Gamma_{m\ell}}{2} 
    \Big(
    F_{m\ell}
        -f^1_{m\ell} - f^M_{m\ell} +2(M+1)
    \Big),
\end{equation}
where we define a series $F_{m\ell}$ as 
\begin{equation} \label{eqn:domain_wall_cs}
    F_{m\ell} \equiv \sum^{M-1}_{i=1} 
        f^i_{m\ell}f^{i+1}_{m\ell}
        -f^i_{m\ell} - f^{i+1}_{m\ell}
\end{equation}


\newpage

\end{document}